\documentclass{article}

\usepackage[nonatbib, preprint]{neurips_2024}

\usepackage[utf8]{inputenc} % allow utf-8 input
\usepackage[T1]{fontenc}    % use 8-bit T1 fonts
\usepackage{hyperref}       % hyperlinks
\usepackage{url}            % simple URL typesetting
\usepackage{booktabs}       % professional-quality tables
\usepackage{amsfonts}       % blackboard math symbols
\usepackage{nicefrac}       % compact symbols for 1/2, etc.
\usepackage{microtype}      % microtypography
\usepackage{xcolor}         % colors

\usepackage{amsmath,amsfonts}
\usepackage{caption}
\usepackage{subcaption}
\usepackage{graphicx} % subcaption for subfigure environment
\usepackage{array}
\usepackage{multirow}
\usepackage{wrapfig}
\usepackage{algorithm}
\usepackage{algorithmic}
\usepackage[capitalize,noabbrev]{cleveref}

\title{Knowledge-enhanced Relation Graph and Task Sampling for Few-shot Molecular Property Prediction}

\author{
Zeyu Wang\textsuperscript{1}
\quad Tianyi Jiang\textsuperscript{1}
\quad Yao Lu\textsuperscript{1}
\quad Xiaoze Bao\textsuperscript{2}
\quad Shanqing Yu\textsuperscript{1}
\\
\quad \textbf{Bin Wei}\textsuperscript{2}
\quad \textbf{Qi Xuan}\textsuperscript{1*}
\\
\textsuperscript{1}{Institute of Cyberspace Security, Zhejiang University of Technology}
\\
\textsuperscript{2}{the College of Pharmaceutical Science \& Collaborative Innovation Center of Yangtze River}\\ 
{Delta Region Green Pharmaceuticals, Zhejiang University of Technology}
\\
\texttt{vencent\_wang@outlook.com, xuanqi@zjut.edu.cn}
}

\begin{document}

\maketitle

\begin{abstract}
Recently, few-shot molecular property prediction (FSMPP) has garnered increasing attention. Despite impressive breakthroughs achieved by existing methods, they often overlook the inherent many-to-many relationships between molecules and properties, which limits their performance. For instance, similar substructures of molecules can inspire the exploration of new compounds. Additionally, the relationships between properties can be quantified, with high-related properties providing more information in exploring the target property than those low-related. To this end, this paper proposes a novel meta-learning FSMPP framework (KRGTS), which comprises the \underline{\textbf{K}}nowledge-enhanced \underline{\textbf{R}}elation \underline{\textbf{G}}raph module and the \underline{\textbf{T}}ask \underline{\textbf{S}}ampling module. The knowledge-enhanced relation graph module constructs the molecule-property multi-relation graph (MPMRG) to capture the many-to-many relationships between molecules and properties. The task sampling module includes a meta-training task sampler and an auxiliary task sampler, responsible for scheduling the meta-training process and sampling high-related auxiliary tasks, respectively, thereby achieving efficient meta-knowledge learning and reducing noise introduction. Empirically, extensive experiments on five datasets demonstrate the superiority of KRGTS over a variety of state-of-the-art methods. The code is available in \url{https://github.com/Vencent-Won/KRGTS-public}.
\end{abstract}

\section{Introduction} \label{sec: introduction}
Molecular property prediction (MPP), aiming to predict the physicochemical properties and biological activity of molecules, is a crucial task in drug discovery. In the past, MPP is expensive, time-consuming, and risky by wet lab experiments. With the advancement of machine learning, lots of MPP models have been developed to ameliorate this situation. For example, deep learning methods showed great potential, such as~\cite{SMILESBERT, KPGT, ASGN} model molecules as text or graph data to explore properties with language models or graph neural networks. However, as the data dependency, limited annotated data hinders MPP models in practical applications. It is essentially a few-shot learning problem. 

Few-shot learning refers to learning from limited samples with supervised information~\cite{song2023comprehensive,li2023libfewshot}. Although the few-shot learning methods are widely studied in computer vision and natural language processing~\cite{lifchitz2019dense, li2023deep}, it is still under-explored in MPP. In recent years, several few-shot learning methods, as are listed in~\cref{tab: FSMPP comparison}, have been introduced to MPP. Specifically, IterRefLSTM~\cite{altae2017low} is a few-shot learning framework that combines the iterative refinement long short-term memory and graph convolution network. Meta-MGNN~\cite{guo2021few} utilizes the self-supervised atom-bond prediction tasks and self-attentive property weight learning module with episodic training diagram in meta-learning. 
However, these methods overlook the correlation between molecules, meaning that labeled molecules can offer valuable information for unlabeled ones. PAR~\cite{wang2021property} emphasized the importance of relationships between different molecules and proposed the property-aware relation between molecules. HSL-RG~\cite{ju2023few} and PG-DERN~\cite{zhang2024property} constructed molecular relation graphs with molecular representations to capture information from support molecules. Furthermore, GS-Meta~\cite{zhuang2023graph} took account of the correlation of properties and developed a contrastive learning meta-training task sampler and the molecule-property relation graph (MPRG) which consists of auxiliary properties, target properties, and molecules. Although these methods noticed the correlation between the molecules, graph embedding-based similarity of molecules may be sensitive to model initialization and may not be as accurate as direct structure comparisons, such as comparing molecular scaffold topology or functional group types. Also, they overlooked the fine-grained molecular relationships and the quantification of the relationships between molecular properties. As is known to all, one of the key factors influencing molecular properties is the molecular substructure, including scaffolds and functional groups~\cite{jiang2024mix,noy1995chemical,caminade2015key}, which serve as essential indicators in laboratory research. Moreover, as for the correlation between properties, the low-related auxiliary properties generally have fewer contributions to target property prediction than high-related properties, which is also reflected in transfer learning~\cite{zhang2023cross}. And, excessive low-related properties will introduce additional computation, information redundancy and make the model smooth, thereby affecting its generalization.

% \begin{table*}[!t]
%     \caption{The comparison of methods for FSMPP.}
%     \centering
%     \resizebox{1\textwidth}{!}{
%     \begin{tabular}{ccccccc}
%         \toprule[1.5pt] %添加表格头部粗线
%         Methods & IterRefLSTM & Meta-MGNN & PAR & HSL-RG & GS-Meta & KRGTS\\
%         \midrule
%         meta-learning & - & \checkmark & \checkmark & \checkmark & \checkmark & \checkmark \\
%         molecules relation & - & - & property-aware & embedding similarity & embedding similarity & substructure similarity\\
%         properties relation & - & - & - & - & meta-training task sampler & meta-training \& auxiliary task sampler \\
%         molecule-property relation & - & - & - & - & molecule-property relation graph & molecule-property relation graph \\
%         \bottomrule[1.5pt] %添加表格底部粗线
%     \end{tabular}
%     }
% \label{tab: complexity}
% \end{table*}

\begin{table}[!t]
    \centering
    \caption{The comparison of few-shot molecular property prediction methods.}
    \label{tab: FSMPP comparison}
    \resizebox{1\textwidth}{!}{
    \begin{tabular}{cccccc}
        \toprule[1.5pt]
        Methods & meta-learning & molecule-property relation & molecule-molecule relation & meta-training task sampler & auxiliary task sampler \\ 
        \midrule
        IterRefLSTM~\cite{altae2017low} & - & - & - & - & -  \\ 
        Meta-MGNN~\cite{guo2021few} & \checkmark & - & - & random sampling & -  \\ 
        PAR~\cite{wang2021property} & \checkmark & - & property aware relation & random sampling & - \\ 
        HSL-RG~\cite{ju2023few} & \checkmark & - & molecular embedding similarity & random sampling & - \\ 
        PG-DERN~\cite{zhang2024property} & \checkmark & - & molecular embedding similarity & random sampling & - \\ 
        GS-Meta~\cite{zhuang2023graph} & \checkmark & \checkmark & molecular embedding similarity & optimized sampler & random sampling  \\
        % ADKF-IFT & \checkmark & - & - & random sampling & - \\ 
        % MolFeSCue & - & - & - & - & - \\ 
        KRGTS & \checkmark & \checkmark & molecular substructure similarity & optimized sampler with task relationships & optimized sampler \\
        \bottomrule[1.5pt]
    \end{tabular}
    }
\end{table}

To handle these problems, this paper delves into fine-grained relationships of molecules and the qualification of relationships between properties, proposing a novel FSMPP framework that comprises the \textbf{K}nowledge-enhanced \textbf{R}elation \textbf{G}raph module and the \textbf{T}ask \textbf{S}ampling module (KRGTS). Specifically, the knowledge-enhanced relation graph module constructs the molecule-property multi-relation graph (MPMRG) incorporating substructure (scaffold and functional group) similarities of molecules and property information, effectively capturing the many-to-many relationships between molecules and properties. Given the scale of MPMRG, KRGTS samples target-centered subgraphs from MPMRG, comprising a target property, molecules, and auxiliary properties, to train the FSMPP model. The task sampling module comprises a meta-training task sampler and an auxiliary task sampler. Among them, the auxiliary task sampler is designed to select high-related auxiliary properties for the target-centered subgraph. Due to the imbalance of data and the connectivity of different subgraphs, it is important to sample suitable meta-training tasks, aiming to effectively accumulate meta-knowledge. The meta-training task sampler utilizes the properties similarity guidance to schedule the training process by sampling subgraphs with different target tasks and is optimized through subgraph-to-subgraph contrastive learning, minimizing the discrepancy between the same target-centered subgraphs while maximizing the discrepancy between different ones. The contributions are summarized as follows:
\begin{itemize}
    \item This paper utilizes the scaffold similarity and functional group similarity with the property information to construct MPMRG, which could effectively capture the many-to-many correlations between molecules and properties.
    \item This paper first proposes to sample the high-related properties as the auxiliary properties to explore the target molecular property. Correspondingly, this paper develops the task sampler module consisting of a meta-training task sampler and an auxiliary task sampler.
    \item Extensive experiments on FSMPP datasets demonstrate the superiority of KRGTS over various state-of-the-art methods. Furthermore, the ROC-AUC of the Tox21 reaches $87.62\%$. 
\end{itemize}

% \vspace{-2mm}
\section{Related Work}

\subsection{Molecular Property Prediction}
Molecular property prediction refers to predicting the physicochemical properties and biological activity of molecules, which plays an important role in virtual screening, drug design, and property optimization process~\cite{walters2020applications}. The mainstream methods of MPP can be divided as molecular descriptors-based methods, Simplified Molecular Input Line Entry Specification (SMILES) based methods, and graph (2D\&3D) based methods~\cite{walters2020applications, feinberg2018potentialnet}. Among them, molecular descriptors-based methods are traditional methods that explore the relationship between properties and the topology or characteristic description such as molecular fingerprint~\cite{CERETOMASSAGUE201558, FingerOverview}. However, the molecular fingerprint can not capture the multi-scale information and 3D conformation information. With the progression of deep learning, SMILES-based methods and graph-based methods showed tremendous potential. SMILES-based methods generally apply the sequence model to learn the molecular representation~\cite{SMILESBERT}, which can effectively encode the atoms and bonds. While compared to the molecular graph, the SMILES lacks structure information and can not distinguish the stereoisomers. Molecules and graph data are highly compatible as the atoms and bonds can be converted to the nodes and edges. There are lots of graph-based methods that encode the molecules with graph neural networks such as the 2D graph~\cite{ijcai2020p392}, the 3D graph~\cite{fang2022geometry} and multi-view methods~\cite{wang2024multi, li2022geomgcl}. Considering the superiority of graph neural network, KRGTS models molecules as graphs to obtain comprehensive molecular representations.

% \vspace{-2mm}
\subsection{Few-shot Molecular Property Prediction}
Few-shot learning has achieved excellent success in computer vision and natural language processing but is still in its fancy in the field of MPP. Recent efforts have been made to release this situation. \cite{guo2021few} proposed that limited data hinder the deep learning application in MPP and developed Meta-MGNN that consists of molecular self-supervised modules and self-attentive task weights module with meta-learning. Taking the relationships of molecules into account, PAR~\cite{wang2021property} introduced the property-aware molecular relation and an adaptive relation learning module. HSL-RG~\cite{ju2023few} utilized the graph kernel to construct relation graphs, enabling the global communication of molecular structural knowledge across neighboring molecules. GS-Meta~\cite{zhuang2023graph} proposed the gap in the consideration of properties relationships and introduced the meta-training task sampler to achieve the effective learning of meta-knowledge. Although these methods have noticed the importance of the relationships between molecules and properties, they overlook the fine-grain relationships such as the substructures (functional groups and scaffolds) relation of different molecules. The graph neural network and graph kernel methods generally pay attention to the global feature and similarity, which can not effectively capture the local feature. Furthermore, they neglect the impact of the relationships of auxiliary properties and different target properties. Additionally, MTA~\cite{meng2023meta} proposed a task augmentation strategy that constructs new samples with highly relevant motifs. However, different from other data (such as images), the properties of virtual molecules are uncertainty factors. To this end, this paper constructs the MPMRG with the substructure similarity of molecules with the property information to capture the auxiliary information provided by annotated molecules. Moreover, to reasonably utilize the information on auxiliary properties that have different relationships with target properties, this paper introduced the auxiliary task sampler to sample the high-related auxiliary properties.

% \vspace{-2mm}
\section{Preliminaries}\label{sec: notations}
\textbf{Molecule-Property Relation Graph (MPRG).} 
A molecule-property relation graph is defined as a graph $G = (V, T, E, B)$, where $V$ is the molecule set, $T$ is the property set, $E=\{(i,\tau)|i\in V,\tau\in T\}$ denotes the molecular property information, and $B$ is the edge weight set. Also, the edge weight set $B=\{b_{i,\tau}|{(i,\tau)\in E}\}$ indicates the label of the molecule $i$ for property $\tau$, where $b\in\{0,1,2\}$ (0 is ${\rm inactive}$, 1 is ${\rm active}$, 2 is ${\rm unknown}$). Moreover, there is no relation between molecules.

\textbf{Molecule-Property Multi-Relation Graph (MPMRG).}
Compared to the molecule-property graph, the molecule-property multi-relation graph not only encompasses molecular properties information but also incorporates relationships among different molecules. Consequently, the molecule-property multi-relation graph can be regarded as a multiplex graph which is a multi-layer view of a graph $M=(V, T, R, \mathbb{M})$. Here, $V$ and $T$ signify the molecules and properties within the molecule-property relation graph, constituting sets of supra-nodes instantiated separately in each layer of the multiplex network. $R$ represents the set of relations, encompassing connections between diverse molecules and from molecules to properties. These relations in the molecule-property relation graph are the foundation for each layer graph within the multiplex network. $\mathbb{M}=\{G^r\}_{r\in R}$ is the set of layer graphs that include several relation graphs.

\textbf{Problem Definition.}
The FSMPP aims to develop a model with the generalization ability to new tasks through training with only a few annotated data. In line with~\cite{guo2021few}, the FSMPP is conducted on a set of tasks $T$ which corresponds to the $N_t$ molecular properties. Then, the task set can be divided into $T_{train}$ and $T_{test}$, corresponding to the training tasks and testing tasks. Generally, the MPP involves qualitative analysis tasks with two classes. Accordingly, the FSMPP task $\tau$ can be formulated as a 2-way $K$-shot classification task $\{S_\tau, Q_\tau\}$, where $S_\tau=\{(v_i,y_{i,\tau})\}_{i=1}^{2K}$ is a set of labeled support molecules, $Q_\tau=\{(v_i,y_{i,\tau})\}_{i=1}^{N_Q}$ is a set of query molecules.

\begin{figure}[!t]
    \centering
    \includegraphics[width=1\linewidth]{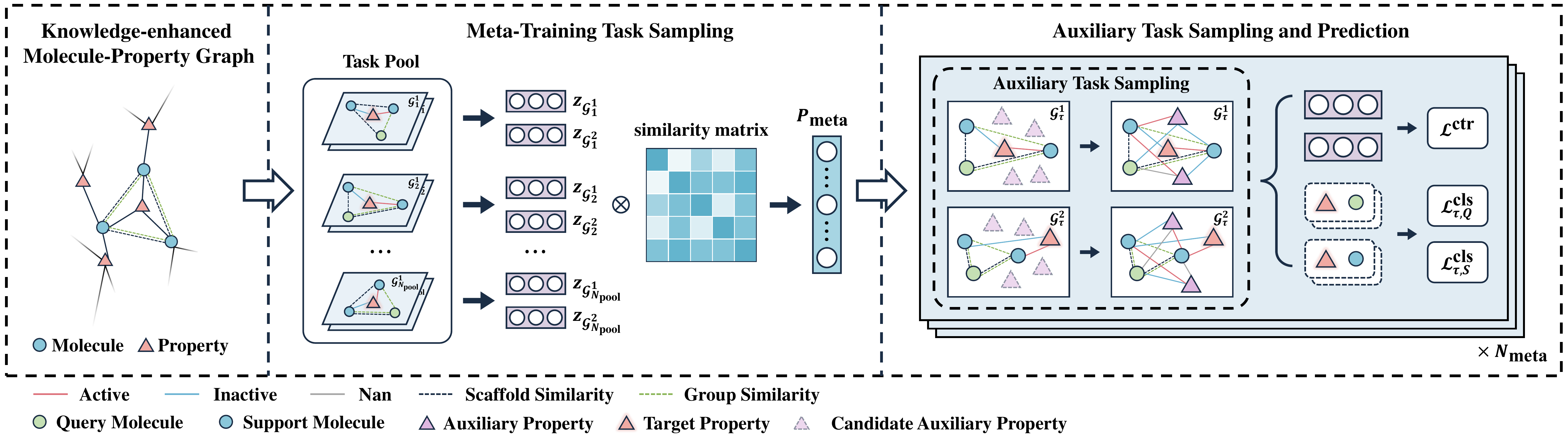}
    \caption{The pipeline of KRGTS in 2-way 1-shot setting.}
    \label{fig: framework}
\end{figure}

\section{Methodology}
% This section commences with a thorough description of the framework of KRGTS. Following that, it introduces the components and implementation of KRGTS in detail.

\subsection{Overview}
\cref{fig: framework} describes the pipeline of KRGTS in 2-way 1-shot setting. Given a FSMPP dataset, one can mainly utilize the substructure similarity of molecules and properties information to construct the knowledge-enhanced molecule-property relation graph, which can be regarded as a MPMRG. Considering the scale of the MPMRG, KRGTS adopts the episodic training paradigm as Meta-MGNN~\cite{guo2021few}. 
%Specifically, KRGTS does not load all meta-training tasks in each epoch but instead uses batches of episodes pairwise subgraphs $\{\mathcal{G}_{\tau}^1, \mathcal{G}_{\tau}^2\}_{\tau=1}^{N_\text{meta}}$ with contrastive learning-based meta-training task sampler, where target-centered subgraph $\mathcal{G}_{\tau}$ consists of support set $S_\tau$ and query set $Q_\tau$. 
Based on the contrastive learning mechanism, one can randomly sample a task pool $\{(\mathcal{G}_{\tau}^1, \mathcal{G}_{\tau}^2)\}_{\tau=1}^{N_\text{pool}}$, consisting of pairwise subgraphs of target tasks, where target-centered subgraph $\mathcal{G}_{\tau}$ consists of support set $S_\tau$ and query set $Q_\tau$. To effectively accumulate meta-knowledge across different meta-training tasks, the meta-training task sampler schedules the training process by sampling a batch of meta-training tasks $\{(\mathcal{G}_{\tau}^1, \mathcal{G}_{\tau}^2)\}_{\tau=1}^{N_\text{meta}}$. Moreover, to capture auxiliary task information and reduce information redundancy, the auxiliary task sampler samples high-related auxiliary tasks (known properties) for target tasks (target properties). Then, one can utilize complete subgraphs to learn comprehensive representations of molecules and properties and apply them to predict the target property. And the algorithm and implementation are listed in~\cref{sec: alg}.

%根据第三节的定义，某可以利用分子及其属性信息构建MPRG。如图2（A）所示，节点代表分子和属性，连边代表分子的属性。虽然该网络与分子属性信息直接相关，但忽略了分子和属性间的多对多关系。HSL_RG,GS-Meta提出通过计算分子表征间的距离作为分子间的相似度来丰富信息。然而该操作不仅忽略了重要的细粒度相似度，并且分子表征受模型初始化影响较大，易导致过拟合。因此，本文提出利用用分子重要子结构（骨架、官能团）相似度来增强分子属性关系网络。如图2（B）所示，KRGTS通过计算分子骨架相似度和官能团相似度来构建MPMRG $M=(V, T, R, \mathbb{M})，其中R$包含了两种分子间子结构相似度和属性信息。而\mathbb{M}=\{G^{Sca}, G^{Gro}, G^{Pro}\}$，其中$G^{Sca}=(V, T, E^{Sca}, B^{Sca})$表示骨架相似层图，$G^{Gro}=(V, T, E^{Gro}, B^{Gro})$表示官能团相似层图，$G^{Pro}=(V, T, E^{Pro}, B^{Pro})$表示属性层。相应地，$B^{Sca}=\{b_{i,j}|(i,j)\in E^{Sca}\}$和$B^{Gro}=\{b_{i,j}|(i,j)\in E^{Gro}\}$表示分子间的相似性, 其中$b\in[0,1]$。

\subsection{Knowledge-enhanced Molecule-Property Relation Graph}
\begin{wrapfigure}{r}{7cm}
    \centering
    \includegraphics[width=1\linewidth]{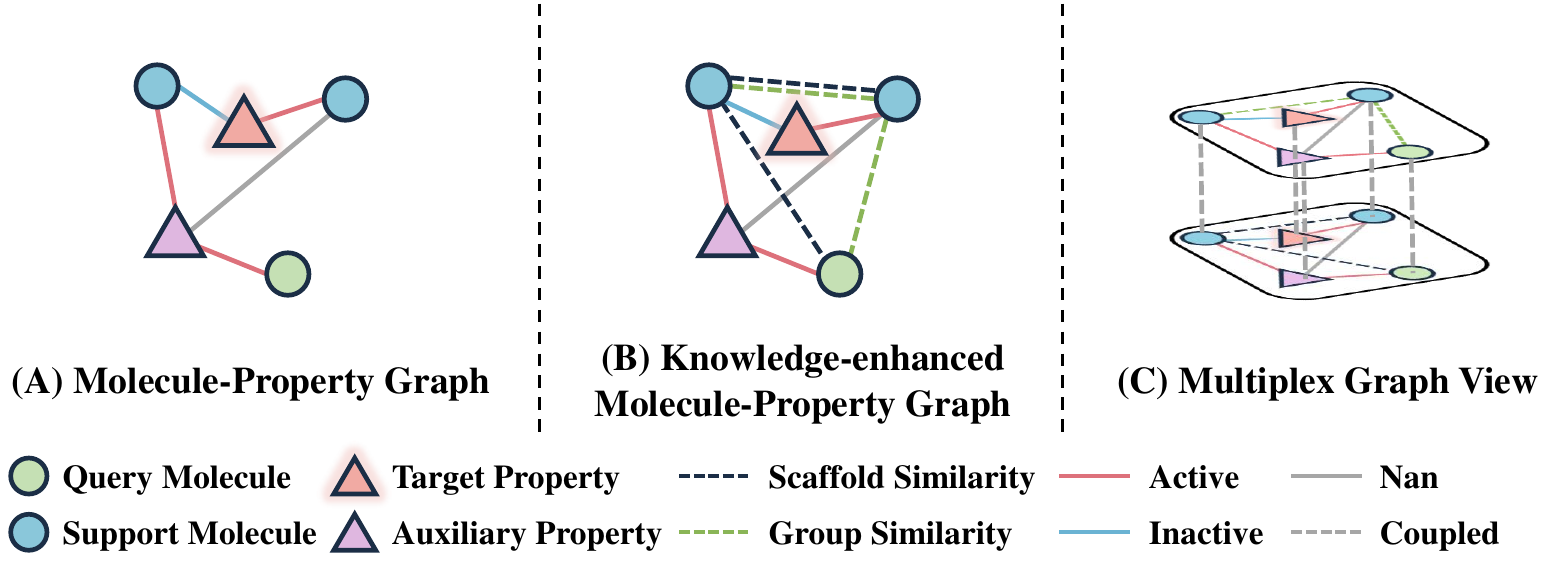}
    \caption{The comparison of molecular-property relation graph and knowledge-enhanced molecule-property graph.}
    \label{fig: subgraph}
\end{wrapfigure}

\textbf{Relation Graph Construction.}
According to the definition in~\cref{sec: notations}, the MPRG can be constructed using molecules and property information. As shown in~\cref{fig: subgraph} (A), nodes represent molecules and properties, and edges represent molecular properties. Although this network intuitively includes property information, it ignores the many-to-many relationships between molecules and properties. HSL\_RG~\cite{ju2023few} and GS-Meta~\cite{zhuang2023graph} propose to utilize molecular representation similarity as molecular similarities to enrich information, these methods not only ignore important fine-grained similarities but also are heavily influenced by model initialization, which can lead to overfitting. Therefore, this paper proposes to augment the molecular property relation network using the similarity of important substructures (scaffolds and functional groups). As illustrated in Figure 2 (B), KRGTS constructs the knowledge-enhanced molecule-property relation graph, which can be regarded as a MPMRG $M=(V, T, R, \mathbb{M})$, where $R$ contains property information and two types of substructure similarity between molecules, and $\mathbb{M}=\{G^\text{Sca}, G^\text{Gro}, G^\text{Pro}\}$. $G^\text{Sca}=(V, T, E^\text{Sca}, B^\text{Sca})$, $G^\text{Gro}=(V, T, E^\text{Gro}, B^\text{Gro})$, and $G^\text{Pro}=(V, T, E^\text{Pro}, B^\text{Pro})$ represent the scaffold similarity layer, the functional group similarity layer, and the property layer, where $B^\text{Sca}=\{b_{i,j}|(i,j)\in E^\text{Sca}\}$ and $B^\text{Gro}=\{b_{i,j}|(i,j)\in E^\text{Gro}\}$ represent the similarity between molecules, and $b\in[0,1]$. Moreover, KRGTS calculates the scaffold similarity and functional group similarity by molecular fingerprints. RDKit\footnote{https://www.rdkit.org/} is employed to extract the molecular scaffold and functional groups. The scaffold similarity of molecules $i$ and $j$ can be denoted as:
\begin{equation}
    {\rm \mathcal{S}_\text{Sca}}(i,j) = \frac{SF_i\cdot SF_j^T}{2214-(\neg SF_i)\cdot (\neg SF_j)^T},
    \label{eq: molecular scaffold similarity}
\end{equation}
where $\neg$ is the negation operator, $SF_i\in\{0,1\}^{2214}$ is the scaffold fingerprint concatenated by the Morgan fingerprint~\cite{morganfingerprint} with the dimension of 2048 and MACCS keys~\cite{maccskey} with the dimension of 166. Each bit of the $SF_i$ indicates whether a structure exists, so that, can capture important structural features such as the ring in scaffolds. Similarly, KRGTS calculates the functional group similarity:
\begin{equation}
    {\rm \mathcal{S}_\text{Gro}}(i,j) = \frac{GF_i\cdot GF_j^T}{49-(\neg GF_i)\cdot (\neg GF_j)^T},
    \label{eq: molecular group similarity}
\end{equation}
where $GF_i\in\{0,1\}^{49}$ corresponds to the set of 49 functional groups defined in RDKit. Each bit of the $GF_i$ indicates whether a functional group exists. 

\textbf{Relation Subgraph Learning.}
Taking the scale of MPMRG into account, KRGTS utilizes the subgraph sampling mechanism with the episodic meta-learning diagram~\cite{finn2017model}. As shown in~\cref{fig: framework}, each task $\tau$ can be formulated as a target-centered subgraph $\mathcal{G}_{\tau}$ with a target property, a query molecule sampled from the query set $Q_{\tau}$, and $2K$ support molecules in the support set $S_{\tau}$. To utilize the correlations between known properties and the target property, the auxiliary task sampler is devised to sample $N_\text{auxi}$ high-related available properties, which are connected to molecules in the subgraph, to assist in predicting the target properties. Then, the subgraph can be denoted as $\mathcal{G}_{\tau}=(\mathcal{V},\{\tau\cup T_\text{auxi}\}, R, \{G^\text{Sca}_{\tau}, G^\text{Gro}_{\tau}, G^\text{Pro}_{\tau}\})$, where $G^\text{Sca}_{\tau}$ and $G^\text{Gro}_{\tau}$ only preserve the top-$k$ similarity relations for each molecule to reduce noise.
Moreover, due to the diversity of relation types and the heterogeneity in the subgraph, traditional relation neural networks~\cite{schlichtkrull2018modeling} cannot be directly applied. To address this challenge, KRGTS treats the $\mathcal{G}_{\tau}$ as a multiplex graph, composed of relation-specific layer graphs with super nodes linked across relations. As depicted in~\cref{fig: subgraph} (C), relation-specific layer graphs correspond to the functional group similarity relation layer with property information $\{G_{\tau}^\text{Pro}, G_{\tau}^\text{Gro}\}$ and the molecular scaffold similarity relation layer with property information $\{G_{\tau}^\text{Pro}, G_{\tau}^\text{Sca}\}$. Subsequently, KRGTS independently conducts message-passing on the two layers and concatenates the outputs to obtain the final representation.

Firstly, KRGTS utilizes the graph encoder~\cite{xu2018how}, the Embedding layers~\cite{paszke2019pytorch}, and radial basis functions~\cite{9338432} to initialize the node (molecules and properties nodes) embedding $h_i^0$ and edge embedding $h_{i,j}$ (details are in~\cref{Details of KRGTS}). For each relation layer, node embeddings are updated as follows:
\begin{equation}
    h_i^l = {\rm GNN}(h_i^{l-1}, h_j^{l-1}, h_{i,j}|j\in \mathcal{N}(i)),
    \label{GNN}
\end{equation}
where $h_i^l$ denotes the embedding of node $i$ after the $l$-th iteration, $\mathcal{N}(i)$ denotes the neighbor nodes of node $i$. After $L$ layers aggregation, one can get the node embeddings of two relation layers and concatenate them to get the final node embeddings of $\mathcal{G}_{\tau}$:
\begin{equation}
    z_i = \sigma({\rm MLP}([h_{\text{Sca},i}^L\oplus h_{\text{Gro},i}^L])),
    \label{prediction vector}
\end{equation}
where $z_i\in\mathbb{R}^d$ denotes the representation of node $i$, $d$ is the dimension of subgraph representation, and $\oplus$ is the concatenation operation. Correspondingly, $z_\tau$ is the final embedding of the target task $\tau$ of $\mathcal{G}_\tau$. To highlight the target task in the subgraph, one can get the subgraph embedding $z_{\scriptscriptstyle \mathcal{G}_\tau}$:
\begin{equation}
    z_{\scriptscriptstyle \mathcal{G}_\tau} = z_\tau + \frac{\sum_{i\in\{\mathcal{V}\cup T_\text{auxi}\}}{z_i}}{|\mathcal{G}_\tau|-1}
    \label{eq: subgraph representation}
\end{equation}
where $|\mathcal{G}_\tau|= |\mathcal{V}|+|\{\tau\cup T_\text{auxi}\}|$ denotes the number of nodes in $\mathcal{G}_\tau$. Finally, KRGTS concatenates the embedding of the molecule $z_{i}$ and the property $z_{\tau}$ to predict target property:
\begin{equation}
    \hat{y}_{i,\tau}=f_\text{clr}([z_{i}\oplus z_{\tau}]),
\end{equation}
where $\hat{y}_{i,\tau}\in \mathbb{R}^1$ is the prediction, and $f_\text{clr}$ is a classifier consisting of two layers ${\rm MLP}$. Therefore, the loss of support molecules is defined as:
\begin{equation}
    \mathcal{L}^\text{cls}_{\tau,S} = -\sum_{v_i\in S_\tau}(y_{i,\tau}\log\hat{y}_{i,\tau}+(1-y_{i,\tau})\log(1-\hat{y}_{i,\tau})).
    \label{eq: support loss}
\end{equation}
Similarly, one can get the loss of query set $\mathcal{L}^\text{cls}_{\tau,Q}$.

%随着序列采样策略在元学习中的流行，该训练机制被逐步采纳用于FSMPP。Meta-MGNN，PAR认为元训练任务是平等重要的，并使用固定概率随机采样元任务。而GS-Meta考虑到不同任务间的相关性，设计了基于对比学习优化的采样器。但它只是简单地将目标任务子图和其他任务子图的表征相加来预测目标任务用于训练的概率，无法有效捕捉不同任务间的相关性。基于此，KRGTS通过构建任务间相似度关系网络来捕捉任务间的联系来引导采样。
\subsection{Meta-training Task Sampler} \label{sec: target task sampling}
As the episodic training diagram showed its success in meta-learning, this training mechanism is gradually adopted for FSMPP. Meta-MGNN~\cite{guo2021few} and PAR~\cite{wang2021property} considered meta-training tasks to be equally important and randomly sample meta-training tasks with a fixed probability. On the other hand, GS-Meta designed a meta-training task sampler with reinforcement learning and introduced contrastive learning to augment the task representation learning designs. However, all of them failed to capture the inherent correlation between different tasks. To address these challenges, KRGTS proposes the meta-training task sampler based on two aspects: 1) Each target-centered subgraph with different molecules can be regarded as different views of the target property, so that, should minimize their semantic discrepancy. Correspondingly, the semantics of target-centered subgraphs with different target properties should be enlarged. 2) Capturing the relationships between tasks can guide the sampler to sample meta-training tasks and optimize the meta-knowledge accumulation.

Firstly, according to the setting of contrastive learning, one can randomly sample a task pool $\{(\mathcal{G}_{\tau}^1, \mathcal{G}_{\tau}^2)\}_{\tau=1}^{N_\text{pool}}$ for sampling, where $\mathcal{G}_{\tau}^1, \mathcal{G}_{\tau}^2$ represents the two views of target task $\tau$. Each $\mathcal{G}_{\tau}$ consists of a target task, a query molecule, and 2$K$ support molecules, i.e., $\mathcal{G}_{\tau}$ contains 2$K$+2 nodes. One can get the subgraph representations $Z_\text{pool}\in\mathbb{R}^{2*N_\text{pool}\times d}$ with~\cref{eq: subgraph representation} and learn a similarity matrix $A_\text{pool}\in \mathbb{R}^{2*N_\text{pool}\times 2*N_\text{pool}}$ to model the relationships of tasks in the task pool, $A_\text{pool}(\tau_1,\tau_2)={\rm \mathcal{S}_{cos}}(z_{\scriptscriptstyle \mathcal{G}_{\tau_{_1}}},z_{\scriptscriptstyle \mathcal{G}_{\tau_{_2}}})$, ${\rm \mathcal{S}_{cos}}$ is the cosine similarity function. Then, one can perform message propagation to enrich the task representations:
%An aggregation operation, which can enhance the information interaction of tasks, is applied to the task-level graph:
\begin{equation}
    Z_\text{pool}^{'} = {\rm ReLU}(A_\text{pool}Z_\text{pool}W_\text{pool}),
\end{equation}
where $W_\text{pool}\in \mathbb{R}^{d\times d}$ is a learnable parameters matrix. Subsequently, one can take the updated representation $z'_{\scriptscriptstyle \mathcal{G}_\tau}\in Z_\text{pool}^{'}$ as input to get the sampling probability $P_{\text{meta},\mathcal{G}_\tau}$:
\begin{equation}
    P_{\text{meta},\mathcal{G}_{\tau}} = f_\text{meta}(z'_{\scriptscriptstyle \mathcal{G}_\tau}),
\end{equation}
where $f_\text{meta}$ consists of $L_\text{meta}$ layers ${\rm MLP}$. The sampling probability of $\tau$ is computed as $(P_{\text{meta},\mathcal{G}_\tau^1}+P_{\text{meta},\mathcal{G}_\tau^2})/2$. KRGTS selects $N_\text{meta}$ meta-training tasks with sampling probabilities for each epoch. Moreover, the NT-Xent loss~\cite{hjelm2018learning} is utilized for subgraph-to-subgraph contrastive learning:
\begin{equation}
    \mathcal{L}^\text{ctr} = \frac{1}{N_\text{meta}}\sum_{\tau=1}^{N_\text{meta}}{-\log{\frac{e^{{\rm \mathcal{S}_{cos}}(z_{\scriptscriptstyle \mathcal{G}_{\tau}^1},z_{\scriptscriptstyle \mathcal{G}_{\tau}^2})/t}}{\sum_{\tau'=1,\tau'\neq \tau}^{N_\text{meta}}{e^{{\rm \mathcal{S}_{cos}}(z_{\scriptscriptstyle \mathcal{G}_{\tau'}^1},z_{\scriptscriptstyle \mathcal{G}_{\tau'}^2})/t}}}}},
    \label{eq: contrastive loss}
\end{equation}
where $t$ is the temperature parameter. To effectively accumulate meta-knowledge, KRGTS takes the contrastive loss as the reward $R_\text{meta}$ for optimizing:
\begin{equation}
    \phi \leftarrow \phi + lr_\text{meta}\nabla_{\phi} P_\text{meta}(R_\text{meta}-b_\text{meta}),
    \label{eq: meta sampler update}
\end{equation}
where $\phi$ denotes parameters of the meta-training task sampler module, $lr_\text{meta}$ is the learning rate of the auxiliary tasks sampler, $P_\text{meta}$ is the sampling probability of selected meta-training tasks, and $b_\text{meta}$ is the moving average of reward.

\subsection{Auxiliary Task Sampler}
Task relevance is not only an important factor in meta-training task scheduling but also plays a pivotal role in relation subgraphs construction. As a crucial source of information within relation subgraphs, the information carried by high-related auxiliary tasks is often richer than that from low-related tasks. The low-related auxiliary tasks may introduce extra noise and result in information redundancy. To this end, KRGTS develops a result-oriented auxiliary task sampler with policy gradient~\cite{williams1992simple}.

Firstly, for each target property $\tau$, the candidate auxiliary properties set is $T_\text{can}=T_\text{train}\setminus \tau$. One can get the embedding of the target property subgraph $z_{\scriptscriptstyle \mathcal{G}_\tau}\in \mathbb{R}^{d}$ and the subgraph embeddings of all candidate auxiliary properties $Z_\text{can}\in \mathbb{R}^{|T_\text{can}|\times d}$. Subsequently, the sampling probability of auxiliary task $\tau_\text{can}\in T_\text{can}$ is computed as:
\begin{equation}
    P_{auxi, \tau_\text{can}} = f_\psi([z_{\scriptscriptstyle \mathcal{G}_{\tau}}\oplus z_{\scriptscriptstyle \mathcal{G}_{\tau_\text{can}}}]),
\end{equation}
where $f_\psi$ consists of $L_\text{auxi}$ layers ${\rm MLP}$ with parameter $\psi$. Furthermore, for each target task $\tau$, KRGTS selects top $N_\text{auxi}$ auxiliary tasks according to the auxiliary tasks sampling probability. Therefore, each $\mathcal{G}_{\tau}$ consists of a target task, a query molecule, 2$K$ support molecules, and $N_\text{auxi}$ auxiliary tasks, i.e., $\mathcal{G}_{\tau}$ contains 2$K$+2+$N_\text{auxi}$ nodes. According to the objective of the auxiliary sampling, it is intuitive to take the value of query loss $\mathcal{L}_{\tau, Q}^\text{cls}$ as reward $R_\text{auxi}$:
\begin{equation}
    \psi \leftarrow \psi + lr_\text{auxi}\nabla_{\psi} P_\text{auxi}(R_\text{auxi}-b_\text{auxi}),
    \label{eq: auxi sampler update}
\end{equation}
where $lr_\text{auxi}$ is the learning rate of the auxiliary tasks sampler, $P_\text{auxi}$ is the sampling probability of selected auxiliary tasks, and $b_\text{auxi}$ is the moving average of reward.

\section{Experiments}
In this section, extensive experiments are conducted on five well-known benchmark datasets to evaluate the effectiveness of KRGTS with the 10-shot and 1-shot settings. Also, the auxiliary task number experiments, task relevance experiments, and the ablation study are conducted to ensure the effects of each module in KRGTS. Moreover, the details of experiments and more experiment results are listed in \cref{sec: experiments details}. 

\begin{table*}[!t]
    \caption{ROC-AUC of few-shot molecular property prediction. The best is marked with boldface and the sub-optimal is with underline.}
    \centering
    \resizebox{1\columnwidth}{!}{
    \begin{tabular}{ccccccccccc}
    \toprule[1.5pt]
    \multirow{2}{*}{Method} & \multicolumn{2}{c}{Tox21} & \multicolumn{2}{c}{SIDER} & \multicolumn{2}{c}{MUV} & \multicolumn{2}{c}{ToxCast} & \multicolumn{2}{c}{PCBA} \\
    \cmidrule{2-11}
    & 10-shot & 1-shot & 10-shot & 1-shot & 10-shot & 1-shot & 10-shot & 1-shot & 10-shot & 1-shot\\
    \midrule
    Siamese & 80.40$_{(0.35)}$ & 65.00$_{(1.58)}$ & 71.10$_{(4.32)}$ & 51.43$_{(3.31)}$ & 59.96$_{(5.13)}$ & 50.00$_{(0.17)}$ & - & - & - & - \\
    ProtoNet & 74.98$_{(0.32)}$ & 65.58$_{(1.72)}$ & 64.54$_{(0.89)}$ & 57.50$_{(2.34)}$ & 65.88$_{(4.11)}$ & 58.31$_{(3.18)}$ & 63.70$_{(1.26)}$ & 56.36$_{(1.54)}$ & 64.93$_{(1.94)}$ & 55.79$_{(1.45)}$ \\
    MAML & 80.21$_{(0.24)}$ & 75.74$_{(0.48)}$ & 70.43$_{(0.76)}$ & 67.81$_{(1.12)}$ & 63.90$_{(2.28)}$ & 60.51$_{(3.12)}$ & 66.79$_{(0.85)}$ & 65.97$_{(5.04)}$ & 66.22$_{(1.31)}$ & 62.04$_{(1.73)}$ \\
    TPN & 76.05$_{(0.24)}$ & 60.16$_{(1.18)}$ & 67.84$_{(0.95)}$ & 62.90$_{(1.38)}$ & 65.22$_{(5.82)}$ & 50.00$_{(0.51)}$ & 62.74$_{(1.45)}$ & 50.01$_{(0.05)}$ & - & - \\
    EGNN & 81.21$_{(0.16)}$ & 79.44$_{(0.22)}$ & 72.87$_{(0.73)}$ & 70.79$_{(0.95)}$ & 65.20$_{(2.08)}$ & 62.18$_{(1.76)}$ & 63.65$_{(1.57)}$ & 61.02$_{(1.94)}$ & 69.92$_{(1.85)}$ & 62.14$_{(1.58)}$ \\
    IterRefLSTM & 81.10$_{(0.17)}$ & 80.97$_{(0.10)}$ & 69.63$_{(0.31)}$ & 71.73$_{(0.14)}$ & 49.56$_{(5.12)}$ & 48.54$_{(3.12)}$ & - & - & - & - \\
    % MHNfs & - & - & - & - & - & - & - & - \\
    PAR & 82.06$_{(0.12)}$ & 80.46$_{(0.13)}$ & 74.68$_{(0.31)}$ & 71.87$_{(0.48)}$ & 66.48$_{(2.12)}$ & 64.12$_{(1.18)}$ & 69.72$_{(1.63)}$ & 67.28$_{(2.90)}$ & 70.05$_{(0.94)}$ & 67.77$_{(1.04)}$ \\
    % HiMPP & 84.26$_{(0.18)}$ & 82.81$_{(0.12)}$ & 82.31$_{(0.21)}$ & 77.90$_{(0.39)}$ & \textbf{71.30$_{(1.53)}$} & 65.66$_{(2.24)}$ & 71.15$_{(1.09)}$ & 69.84$_{(1.20)}$ \\
    GS-Meta & \underline{85.85$_{(0.26)}$} & \underline{84.32$_{(0.89)}$} & \underline{83.72$_{(0.64)}$} & \underline{82.84$_{(0.67)}$} & 67.11$_{(1.95)}$ & 64.70$_{(2.88)}$ & \underline{81.55$_{(0.19)}$} & \underline{80.03$_{(0.26)}$} & \underline{72.16$_{(0.71)}$} & \underline{70.03$_{(1.56)}$} \\
    HSL-RG$^-$ & 80.95$_{(0.26)}$ & 79.65$_{(0.22)}$ & 74.66$_{(0.52)}$ & 71.77$_{(0.79)}$ & 70.38$_{(1.35)}$ & \underline{67.22$_{(1.56)}$} & 70.70$_{(1.02)}$ & 70.06$_{(1.05)}$ & - & - \\
    ADKF-IFT & 82.42$_{(0.60)}$ & 77.94$_{(0.91)}$ & 67.72$_{(1.21)}$ & 58.69$_{(1.44)}$ & \textbf{98.18$_{(3.05)}$} & 67.04$_{(4.86)}$ & 72.07$_{(0.81)}$ & 67.50$_{(1.23)}$ & - & - \\
    \midrule
    \textbf{KRGTS} & \textbf{87.19$_{(0.11)}$} & \textbf{86.49$_{(0.18)}$} & \textbf{84.83$_{(0.15)}$} & \textbf{84.74$_{(0.20)}$} & \underline{72.63$_{(1.99)}$} & \textbf{68.93$_{(0.65)}$} & \textbf{82.61$_{(0.52)}$} & \textbf{81.65$_{(0.15)}$} & \textbf{76.61$_{(0.80)}$} & \textbf{74.15$_{(1.10)}$} \\
    \bottomrule[1.5pt]
    \end{tabular}%
    }
    \label{tab: result}
\end{table*}

\begin{table*}[!t]
    \caption{ROC-AUC obtained with a pre-trained GNN encoder. The best is marked with boldface and the sub-optimal is with underline.}
    \centering
    % % \renewcommand{\arraystretch}{1.5}
    \resizebox{1\columnwidth}{!}{
    \begin{tabular}{ccccccccccc}
    \toprule[1.5pt]
    \multirow{2}{*}{Method} & \multicolumn{2}{c}{Tox21} & \multicolumn{2}{c}{SIDER} & \multicolumn{2}{c}{MUV} & \multicolumn{2}{c}{ToxCast} & \multicolumn{2}{c}{PCBA} \\
    \cmidrule{2-11}
    & 10-shot & 1-shot & 10-shot & 1-shot & 10-shot & 1-shot & 10-shot & 1-shot & 10-shot & 1-shot\\
    \midrule
    Pre-GNN & 82.14$_{(0.08)}$ & 81.68$_{(0.09)}$ & 73.96$_{(0.08)}$ & 73.24$_{(0.12)}$ & 67.14$_{(1.58)}$ & 64.51$_{(1.45)}$ & 73.68$_{(0.74)}$ & 72.90$_{(0.84)}$ & - & - \\
    Meta-MGNN & 82.94$_{(0.10)}$ & 82.13$_{(0.13)}$ & 75.43$_{(0.21)}$ & 73.36$_{(0.32)}$ & 68.99$_{(1.84)}$ & 65.54$_{(2.13)}$ & - & -  & - & - \\
    Pre-PAR & 84.93$_{(0.11)}$ & 83.01$_{(0.09)}$ & 78.08$_{(0.16)}$ & 74.46$_{(0.29)}$ & 69.96$_{(1.37)}$ & 66.94$_{(1.12)}$ & 75.12$_{(0.84)}$ & 73.63$_{(1.00)}$ & 73.71$_{(0.61)}$ & 72.49$_{(0.61)}$ \\
    % Pre-HiMPP & 86.31$_{(0.18)}$ & 84.39$_{(0.11)}$ & 84.07$_{(0.18)}$ & 80.79$_{(0.25)}$ & \underline{73.38$_{(1.41)}$} & 68.81$_{(1.29)}$ & 76.36$_{(0.88)}$ & 75.20$_{(0.92)}$ \\
    HSL-RG & 85.56$_{(0.28)}$ & 84.09$_{(0.20)}$ & 78.99$_{(0.33)}$ & 77.53$_{(0.41)}$ & 71.26$_{(1.08)}$ & 68.76$_{(1.05)}$ & 76.00$_{(0.81)}$ & 74.40$_{(0.82)}$ & - & - \\
    MolFeSCue & 85.93$_{(0.10)}$ & 82.05$_{(0.11)}$ & 79.08$_{(0.14)}$ & 73.13$_{(0.56)}$ & 72.96$_{(1.18)}$ & 67.32$_{(1.08)}$ & 76.39$_{(1.52)}$ & 74.82$_{(1.39)}$ & - & - \\
    PG-DERN & 85.25$_{(0.29)}$ & 84.12$_{(0.08)}$ & 79.62$_{(0.32)}$ & 77.69$_{(0.38)}$ & 71.65$_{(0.26)}$ & \textbf{69.66$_{(1.02)}$} & 75.21$_{(0.19)}$ & 74.51$_{(0.17)}$ & - & - \\
    Pre-GS-Meta & \underline{86.91$_{(0.41)}$} & \underline{86.46$_{(0.55)}$} & \underline{85.08$_{(0.54)}$} & \underline{84.45$_{(0.26)}$} & 70.18$_{(1.25)}$ & 67.15$_{(2.04)}$ & \underline{83.81$_{(0.16)}$} & \underline{81.57$_{(0.18)}$} & \underline{79.40$_{(0.43)}$} & \underline{78.16$_{(0.47)}$} \\
    Pre-ADKF-IFT & 86.06$_{(0.35)}$ & 80.97$_{(0.48)}$ & 70.95$_{(0.60)}$ & 62.16$_{(1.03)}$ & \textbf{95.74$_{(0.37)}$} & 67.25$_{(3.87)}$ & 76.22$_{(0.13)}$ & 71.13$_{(1.15)}$  & - & - \\
    % ATGNN(Pre-PAR) & 86.05$_{(0.15)}$ & 84.12$_{(0.21)}$ & 79.88$_{(0.14)}$ & 75.84$_{(0.16)}$ & 71.48$_{(1.47)}$ & 68.03$_{(1.54)}$ & 76.08$_{(0.52)}$ & 74.37$_{(0.81)}$ \\
    % MTA(Pre-PAR) & 86.69$_{(0.73)}$ & 84.15$_{(0.60)}$ & 79.73$_{(0.88)}$ & 76.53$_{(0.94)}$ & 71.49$_{(1.06)}$ & \underline{70.75$_{(1.15)}$} & 76.27$_{(1.12)}$ & 75.29$_{(0.92)}$ \\
    \midrule
    \textbf{Pre-KRGTS} & \textbf{87.62$_{(0.29)}$} & \textbf{87.54$_{(0.11)}$} & \textbf{85.09$_{(0.31)}$} & \textbf{84.61$_{(0.16)}$} & \underline{74.47$_{(0.82)}$} & \underline{68.69$_{(0.60)}$} & \textbf{84.02$_{(0.10)}$} & \textbf{82.39$_{(0.29)}$} & \textbf{81.59$_{(0.30)}$} & \textbf{81.18$_{(0.17)}$} \\
    \bottomrule[1.5pt]
    \end{tabular}%
    }
    \label{tab: pre-result}
\end{table*}

\subsection{Overall Performance Comparison} \label{sec: Overall Performance Comparison}
Here, following the comparison experiments designed in~\cite{wang2021property,meng2023meta}, this paper conducts both the 2-way 10-shot and 2-way 1-shot MPP on Tox21\footnote{\url{https://tripod.nih.gov/tox21/challenge/}}, SIDER
~\cite{kuhn2016sider}, MUV~\cite{rohrer2009maximum}, ToxCast~\cite{richard2016toxcast} and PCBA~\cite{wang2012pubchem}.~\cref{tab: result} and~\cref{tab: pre-result} lists the comparison results based on ROC-AUC and standard deviation of KRGTS against 18 well-known baselines. Based on the observations in \cref{tab: result} and \cref{tab: pre-result} among meta-learning methods, graph neural network methods, and other FSMPP methods, it is evident that KRGTS consistently achieves the best performance in terms of ROC-AUC and its standard deviation on Tox21, SIDER, ToxCast, and PCBA. Specifically, KRGTS outperforms the sub-optimal baselines by $6.17\%$ on PCBA, and particularly achieves $87.62\%$ ROC-AUC on Tox21 in the 10-shot setting. This superior performance can be credited to its ability to capture the many-to-many relationships between molecules and properties. On the MUV dataset with the 10-shot setting, KRGTS performs less effectively than ADKF-IFT. According to the dataset statistics in~\cref{sec: experiments details}, this phenomenon may be attributed to the sparsity of property labels for MUV molecules, which impacts the model performance. Moreover, the performance of KRGTS on other datasets and the MUV dataset in the 1-shot setting is far superior to ADKF-IFT.

\begin{figure}[!t]
    \centering
    \includegraphics[width=1\linewidth]{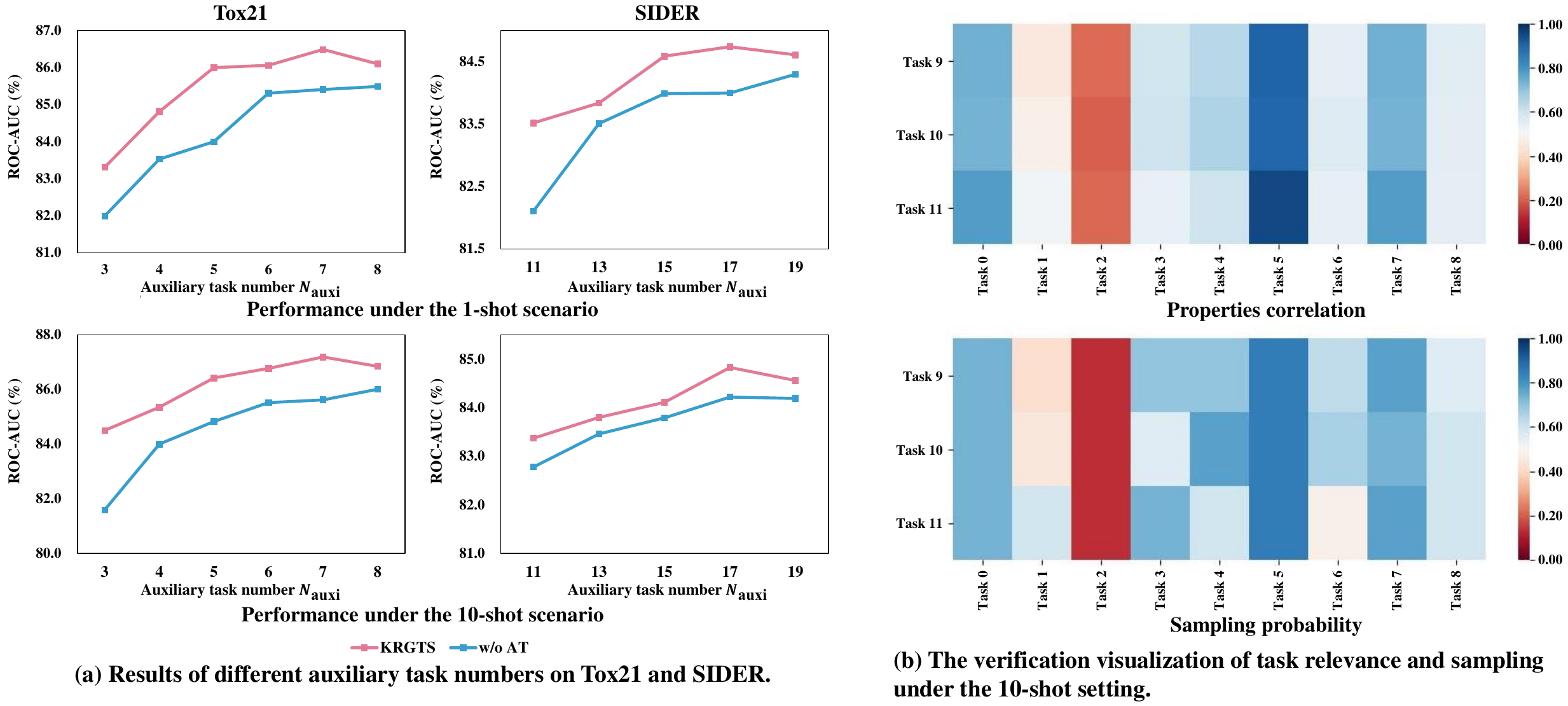}
    \caption{Experiments on the auxiliary task sampler.}
    \label{fig: two graphs}
\end{figure}

% \begin{figure}[!t]
%     \centering
%     \begin{subfigure}[b]{0.48\textwidth}
%         \centering
%         \includegraphics[width=\textwidth]{Styles/graph/parameter.pdf}
%         \caption{Results of different auxiliary task numbers on Tox21.}
%         \label{fig: parameter}
%     \end{subfigure}%
%     \hfill
%     \begin{subfigure}[b]{0.48\textwidth}
%         \centering
%         \includegraphics[width=\textwidth]{Styles/graph/sample probability.pdf}
%         \caption{The verification visualization of task relevance and sampling under the 10-shot setting.}
%         \label{fig: sampler visualization}
%     \end{subfigure}%
%     \caption{Experiments on the auxiliary task sampler.}
%     \label{fig: two graphs}
% \end{figure}

\subsection{Analysis of Auxiliary Task Sampler} \label{sec: Parameter Analysis}
%为进一步验证辅助任务以及辅助任务采样器在目标任务预测中作用，本文基于两个问题展开探究：1）辅助任务数量是越多越好吗？；2）辅助任务采样器是否能帮助模型在不同辅助任务数量下均取得最优的性能。基于此，本文在Tox21和SIDER数据集上基于多个辅助任务数量的设定对比了KRGTS的辅助任务采样器和随机采样器。实验结果如图3（A）所示。总体来说，随着辅助任务数量的增加，模型性能呈现上升趋势，但当辅助任务数量超过某个值时会出现下降趋势。同时，任意辅助任务数量设定下，KRGTS的性能往往优于随机采样。基于以上现象易得出一下结论：1）辅助任务数量并非越多越好，过多的辅助任务会引入大量噪声；2）KRGTS对于任务间关系的捕捉有效地提升了模型的预测性能。
To delve into the practical effect of auxiliary tasks and the auxiliary task sampler in FSMPP, this study conducts detailed investigations around two questions: \textbf{(RQ1)} Should the number of auxiliary tasks be maximized? \textbf{(RQ2)} Can the auxiliary task sampler consistently boost model performance under different numbers of auxiliary tasks? With this aim, this subsection compares the performance of KRGTS's auxiliary task sampler with the random sampler based on various numbers of auxiliary task settings on the Tox21 and SIDER datasets. 

As illustrated in~\cref{fig: two graphs} (a), the experiment results reveal several important findings: Firstly, as the number of auxiliary tasks gradually increases, the overall model performance shows an upward trend. However, when the number of auxiliary tasks exceeds a certain threshold, the model performance begins to decline; secondly, regardless of how the number of auxiliary tasks is set, the KRGTS auxiliary task sampler consistently demonstrates superior performance compared to the random sampler. Based on these experiment phenomenons, it is easy to conclude that: 1) There exists a balance point for the number of auxiliary tasks, beyond which excessive auxiliary tasks may introduce noise and lead to a decrease in model performance; 2) KRGTS effectively captures relationships between tasks and improves prediction performance through intelligent sampling strategies.
% Here, to explore the impact of auxiliary tasks on target task prediction, parameter experiments based on the number of auxiliary tasks and ablation experiments on auxiliary tasks sampling were conducted on Tox21 and SIDER. The results are presented in~\cref{fig: parameter}, where the without auxiliary task sampling\textbf{(w/o AT)} means the random sampling and the number of auxiliary tasks remains consistent during both training and testing. Since the 9 refers to using all candidate auxiliary tasks in training and testing, the results of KRGTS and w/o AT are consistent. 

% From the observations in~\cref{fig: parameter}, it can be concluded that: (1) On the one hand, comparing KRGTS with w/o AT reveals that when the number of auxiliary tasks is limited, optimization-based sampling often outperforms random sampling. This also confirms the existence of correlations between different molecular property tasks, indicating that auxiliary tasks highly correlated with the target property are more beneficial for predicting the target property. (2) On the other hand, the performance curve of KRGTS based on the number of auxiliary tasks reveals that both excessive and few auxiliary tasks can negatively affect the prediction of the target task. This also suggests the importance of sampling an appropriate number of auxiliary tasks. Moreover, as random sampling cannot capture task correlations, this phenomenon is not manifested in the experiments using random sampling. 

\subsection{Analysis of Task Relevance} \label{sec: Analysis of Task Relevance}
Capturing task relevance is one of the key characteristics of KRGTS. To verify the task relevance capture ability of the auxiliary task sampler, the relationships of properties and the sampling probability of the auxiliary task sampler on Tox21 under the 2-way 10-shot setting are visualized in~\cref{fig: two graphs} (b). Specifically, the properties correlations are Pearson correlation coefficients calculated by task-centered subgraph embeddings acquired from the model after sufficient iterations, where tasks 9-11 are the test set $T_{test}$ of Tox21, and tasks 0-8 are the training set $T_{train}$. KRGTS speculated that auxiliary properties having high-related scores with the target properties would be more beneficial for the target properties prediction. The sampling probability heatmap corresponds to the output of the auxiliary task sampler when sampling auxiliary tasks from tasks 0-8 for target tasks 9-11. As is shown in~\cref{fig: two graphs}, it is observed that candidate auxiliary tasks having high-related scores regarding the target tasks are always assigned with higher sampling probabilities, demonstrating that KRGTS can effectively capture the correlation of different properties and sample beneficial auxiliary tasks to improve the property prediction performance.

\subsection{Ablation Study} \label{sec: Ablation Study}
In this part, to further analyze the contribution of KRGTS, the ablation study is conducted on Tox21 with the ablation of each component of KRGTS. Specifically, the ablation experiments include: (1) \textbf{w/o S}: without the scaffold similarity relation between molecules, (2) \textbf{w/o G}: without using the functional group similarity relation between molecules, (3) \textbf{w/o S \& G}: without using the relation between molecules, (4) \textbf{w/o MT}: randomly select meta-training tasks, (5) \textbf{w/o AT}: randomly select auxiliary tasks, (6) \textbf{w/o MT \& AT}: randomly select meta-training tasks and auxiliary tasks.

\begin{wraptable}{r}{5cm}
    \caption{Ablation studies on Tox21.}
    \centering
    \resizebox{0.35\columnwidth}{!}{
    \begin{tabular}{cccc}
    \toprule[1.5pt] %添加表格头部粗线
    Method & 10-shot & 1-shot \\
    \midrule
     w/o S & 85.97$_{(0.54)}$ & 85.58$_{(0.25)}$ \\
     w/o G & 85.55$_{(0.20)}$ & 85.58$_{(0.13)}$ \\
     w/o S \& G & 85.36$_{(0.37)}$ & 84.92$_{(0.72)}$ \\
     w/o MT & 86.06$_{(0.22)}$ & 85.43$_{(0.40)}$ \\
     w/o AT & 85.62$_{(0.33)}$ & 85.41$_{(0.65)}$ \\
     w/o MT \& AT & 85.01$_{(0.74)}$ & 85.13$_{(0.57)}$ \\
     \midrule
     \textbf{KRGTS} & \textbf{87.19$_{(0.11)}$} & \textbf{86.49$_{(0.18)}$} \\
    \bottomrule[1.5pt] %添加表格底部粗线
    \end{tabular}
    }
    \label{tab: Ablation}
\end{wraptable}
\cref{tab: Ablation} lists the experimental results of all variants of KRGTS on the Tox21 dataset. Overall, KRGTS significantly outperforms all its variants, which undoubtedly demonstrates the complementarity and indispensability of each module in KRGTS. Specifically, for the knowledge-enhanced molecule-property relation graph module, relying solely on a single similarity relationship leads to a significant drop in performance. The performance degradation is most noticeable when relationships between molecules are not used. Despite this, the variants still maintain competitiveness compared with most baselines, which highlights the contribution of fine-grained similarity relationships between molecules. 
%The above phenomenon can be attributed to two factors: 1) Molecular scaffolds and functional groups, as crucial components of molecules, jointly influence the properties of molecules; 2) The limitations of single substructure information may result in information bias. 
Furthermore, variants based on samplers exhibit decreased performance, emphasizing the importance of capturing property relationships in property prediction. Among them, the auxiliary task sampler has a more significant impact than the meta-training task sampler. This phenomenon may be attributed to the fact that the auxiliary task sampler is responsible for sampling auxiliary tasks for the target task, which is directly related to the subgraph.

% \subsection{Time Complexity}

% \begin{figure}[!t]
%     \centering
%     \includegraphics[width=0.8\linewidth]{Styles/graph/sample probability.pdf}
%     \label{fig: sampler visualization}
% \end{figure}

\section{Conclusion} \label{sec: conclusion}
This paper proposes a novel FSMPP framework KRGTS consisting of a knowledge-enhanced molecule-property relation graph learning module and a task sampling module. Specifically, KRGTS constructs the knowledge-enhanced molecule-property relation graph based on the relation of molecular substructure (scaffold similarity and functional group similarity) to capture the information provided by annotated molecules. To effectively utilize the correlation between tasks, KRGTS designs the meta-training task sampler to schedule the training process for better meta-knowledge accumulation and the auxiliary task sampler to select auxiliary tasks having high-related scores with target tasks for more effective property prediction. Comprehensive experiments exhibit the superiority of the KRGTS among the state-of-the-art methods. Also, extra experiment results demonstrate the effectiveness of KRGTS and the contribution of the molecular-property relation graph, the meta-training task sampler as well as the auxiliary task sampler.

However, since molecules are highly complex entities, how to effectively capture the relationships between molecules in the molecule-property relation graph still needs to be optimized. When faced with a considerable number of candidate auxiliary properties, designing models to adaptively sample an appropriate number of auxiliary properties for the target task remains challenging. Additionally, besides qualitative MPP, quantitative analysis tasks are also crucial. Therefore, in the future, we plan to explore FSMPP from these several directions.

% \section*{Acknowledgments}
% This should be a simple paragraph before the References to thank those individuals and institutions who have supported your work on this article.
% \input{neurips_2024.bbl}
\bibliographystyle{unsrt}
\bibliography{neurips_2024}

\clearpage

\appendix
\section{List of Abbreviations}\label{sec: Abbreviation}
Here we list the abbreviations of concepts used in this paper.
\vspace{-5mm}
\begin{table}[!h]
    \caption{List of Abbreviations.}
    \centering
    \renewcommand{\arraystretch}{1}
    \resizebox{0.6\columnwidth}{!}{
    \begin{tabular}{cc}
    \toprule[1.5pt] %添加表格头部粗线
    Abbreviation & Meaning \\ 
    \midrule
    MPP & Molecular Property Prediction\\
    FSMPP & Few-shot Molecular Property Prediction\\
    MPRG & Molecule-property Relation Graph\\
    MPMRG & Molecule-property Multi-relation Graph\\
    \bottomrule[1.5pt] %添加表格底部粗线
    \end{tabular}
    }
    \label{tab: Abbreviation}
\end{table}

\section{Alogrithm of KRGTS}\label{sec: alg}
\begin{algorithm}[!h]
  \caption{Alogrithm of KRGTS}
  \label{alg: KRGTS}
  \begin{algorithmic}
  \REQUIRE Konwledge-enhanced molecule-property relation graph $M$
  \ENSURE Relation Subgraph Learning Module $f_{\theta}$, Meta-training task sampler $f_{\phi}$, Auxiliary task sampler $f_{\psi}$
  \FOR{epoch=1,$\cdots$, epochs}
      \STATE Sample the task pool $\{(\mathcal{G}_t^1, \mathcal{G}_t^2)\}_{t=1}^{N_\text{pool}}$ from $M$.
      \STATE Sample $\{(\mathcal{G}_t^1, \mathcal{G}_t^2)\}_{t=1}^{N_\text{meta}}$ with $f_{\phi}$.
      \FOR{$\tau=1,\cdots,N_\text{meta}$}
        \STATE Sample $N_\text{auxi}$ auxiliary properties with $f_{\psi}$ for $\mathcal{G}_{\tau}^1, \mathcal{G}_{\tau}^2$.
        \STATE Calculate the classification loss of the support set by~\cref{eq: support loss} on $\mathcal{G}_\tau^1, \mathcal{G}_\tau^2$.
        \STATE Inner-update the relation subgraph learning module parameters by \cref{eq: inner update}.
        \STATE Calculate the classification loss of the query molecules as~\cref{eq: support loss} on $\mathcal{G}_\tau^1, \mathcal{G}_\tau^2$.
      \ENDFOR
      \STATE Calculate contrastive loss by \cref{eq: contrastive loss}
      \STATE Outer-update the relation subgraph learning module parameters by \cref{eq: outer update}. 
      \STATE Update the meta-training task sampler parameters by \cref{eq: meta sampler update}.
      \STATE Update the auxiliary task sampler parameters by \cref{eq: auxi sampler update}.
  \ENDFOR
  \end{algorithmic}
\end{algorithm}
As is described in~\cref{alg: KRGTS}, the training process involves two loops with the episodic training diagram. Firstly, the meta-training task sampler is utilized to sample $N_\text{meta}$ meta-training task from the task pool. For each meta-training task training or in the inner loop, the auxiliary task sampler is used to sample $N_\text{auxi}$ auxiliary properties to assist in target property prediction. The loss of support set will be used to update the relation subgraph learning module parameters $\theta$ in the inner loop:
\begin{equation}
    \theta \leftarrow \theta-lr_{inner}\nabla_{\theta}\mathcal{L}^{cls}_{t,S}
    \label{eq: inner update}
\end{equation}
where $lr_{inner}$ is the learning rate in inner loop. After the training of $N_\text{meta}$ meta-training tasks, relation subgraph learning module parameters are updated using the query classification loss and contrastive loss:
\begin{equation}
    \theta \leftarrow \theta-lr_{outer}\nabla_{\theta}(\lambda_{ctr}\mathcal{L}^{ctr}+\sum_{t=1}^{N_\text{meta}}\mathcal{L}^{cls}_{t,Q})
    \label{eq: outer update}
\end{equation}
where $lr_{outer}$ is the learning rate across the outer loop, and $\lambda_{ctr}$ is a hyperparameter. Then, the meta-training task sampler and the auxiliary task sampler are updated in order. Among them, considering the diversity of molecular properties, the auxiliary task sampler adopts the batch update to enhance its stability and efficiency.

\section{Relation Subgraph Learning Module} \label{Details of KRGTS}
The embedding of molecules and properties nodes $\mathcal{V}, \mathcal{T}$ can be initialized as:
\begin{equation}
    h_i^0 = 
    \begin{cases}
    f_{\rm mol}(i)& {\rm For\ } i\in \mathcal{V}\\
    {\rm Embedding}(i)& {\rm For\ } i \in \{\tau\cup T_\text{auxi}\}
    \end{cases}
    \label{node embedding}
\end{equation}
where $h_i^0\in\mathbb{R}^{d}$ is the embedding of node $i$ with dimension $d$, $f_{\rm mol}$ is a graph encoder~\cite{xu2018how}, and the $\rm{Embedding}$ is an embedding layer in Pytorch~\cite{Paszke_PyTorch_An_Imperative_2019}. Different from the node embedding, edge embeddings should incorporate the information of relation types and edge weights. Therefore, KRGTS develops encoding methods tailored to the relationship types and corresponding edge weights. Firstly, the relation types $R$ contain three types: molecule-property relation, scaffold similarity molecules relation, and functional group similarity molecules relation. The embedding of each relation type $x_r$ can be randomly initialized with the same length as molecule embedding by the Embedding layer. Since the edge information is different in the corresponding relation graph, KRGTS designs three types of edge weight encoding methods. Among them, the molecule-property edge weights $B^\text{Pro}=\{b_{i,\tau}|{(i,\tau)\in E^\text{Pro}}\}$ contains the label of molecular properties (0 is inactive, 1 is active, 2 is unknown). Similar to the relation type, the embedding of the molecule-property edge weight $x_b$ can be randomly initialized with the same length as molecule embedding by the Embedding layer. While the scaffold similarity molecules edge weights $B^\text{Sca}=\{b_{i,j}|(i,j)\in E^\text{Sca}\}$ and the functional group similarity molecules edge weights $B^\text{Gro}=\{b_{i,j}|(i,j)\in E^\text{Gro}\}$ are not integer type ($b_{i,j} \in [0,1]$), following the previous work~\cite{9338432}, KRGTS adopts to two RBF layers with ${\rm MLP}$ layer to encode diverse molecular similarity:
\begin{equation}
    x_{_{b^{^\text{Sca}}}}={\rm MLP}({\rm RBF}(b^{^\text{Sca}}))
\end{equation}
\begin{equation}
    x_{_{b^{^\text{Gro}}}}={\rm MLP}({\rm RBF}(b^{^\text{Gro}}))
\end{equation}
where the dimension of $x_{\scriptscriptstyle b^{^\text{Sca}}}$ and $x_{\scriptscriptstyle b^{^\text{Gro}}}$ are same as the molecules embeddings. Then, for each edge $(i,j)$, one can concatenate the edge relation type embedding $x_r$ and edge weight embedding $x_b$ to get the embedding $h_{i,j}$:
\begin{equation}
    h_{i,j} = \sigma({\rm MLP}([x_r\oplus x_b]))
\end{equation}
where $\sigma$ is the activation function, $\oplus$ is a concatenation operation, and $(i,j)$ belongs to the $G^r$.

\section{Supplementary for Experiments} \label{sec: experiments details}
The comprehensive experiments are conducted on five well-known FSMPP datasets: Tox21, SIDER, MUV, ToxCast, and PCBA. \cref{tab: dataset} shows the detailed statistics of these benchmark datasets. Regarding the data split, Tox21, SIDER, and MUV datasets followed the split settings provided by~\cite{altae2017low}. For PCBA, the split setting referred to~\cite{zhuang2023graph}. Additionally, due to the sparsity of ToxCast, the dataset was divided into nine sub-datasets as~\cite{zhuang2023graph}. As illustrated in~\cref{tab: toxcast_dataset}, each sub-dataset corresponds to specific properties. 

\begin{table}[!t]
    \caption{Summary of datasets.}
    \centering
    \renewcommand{\arraystretch}{1}
    \resizebox{0.6\columnwidth}{!}{
    \begin{tabular}{cccccc}
    \toprule[1.5pt] %添加表格头部粗线
    Dataset & Tox21 & SIDER & MUV & ToxCast & PCBA \\
    \midrule
    \# Compounds & 7831 & 1427 & 93127 & 8575 & 437929 \\
    \# Tasks & 12 & 27 & 17 & 617 & 128 \\
    \# Train Tasks & 9 & 21 & 12 & 451 & 118 \\
    \# Test Tasks & 3 & 6 & 5 & 158 & 10 \\
    \midrule
    \% Missing Label & 17.05 & 0 & 84.21 & 14.97 & 39.92 \\
    \% Label \textit{active} & 6.24 & 56.76 & 0.31 & 12.60 & 0.84 \\
    \% Label \textit{inactive} & 76.71 & 43.24 & 15.76 & 72.43 & 59.84 \\
    \bottomrule[1.5pt] %添加表格底部粗线
    \end{tabular}
    }
    \label{tab: dataset}
\end{table}

\begin{table*}[!t]
    \caption{Detailed information of sub-datasets on ToxCast dataset.}
    \centering
    \renewcommand{\arraystretch}{1}
    \resizebox{1\columnwidth}{!}{
    \begin{tabular}{cccccccccc}
    \toprule[1.5pt] %添加表格头部粗线
    Sub-dataset & APR & ATG & BSK & CEETOX & CLD & NVS & OT & TOX21 & Tanguay \\
    \midrule
    \# Compounds & 1039 & 3423 & 1445 & 508 & 305 & 2130 & 1782 & 8241 & 1039 \\
    \# Tasks & 43 & 146 & 115 & 14 & 19 & 139 & 15 & 100 & 18 \\
    \# Train Tasks & 33 & 106 & 84 & 10 & 14 & 100 & 11 & 80 & 13 \\
    \# Test Tasks & 10 & 40 & 31 & 4 & 5 & 39 & 4 & 20 & 5 \\
    \midrule
    \% Missing Label & 28.09 & 0.16 & 0 & 1.36 & 0.98 & 92.27 & 2.44 & 8.35 & 1.11 \\
    \% Label \textit{active} & 10.30 & 5.92 & 17.71 & 22.26 & 30.72 & 3.21 & 9.78 & 5.39 & 8.05 \\
    \% Label \textit{inactive} & 61.61 & 93.92 & 82.29 & 76.38 & 68.30 & 4.52 & 87.78 & 86.26 & 90.84 \\
    \bottomrule[1.5pt] %添加表格底部粗线
    \end{tabular}
    }
    \label{tab: toxcast_dataset}
\end{table*}
\subsection{Details of Baselines} \label{Details of Baselines}
To conduct a comprehensive comparison, this paper adopts three types of baselines consisting of meta-learning methods, graph neural network methods, and other FSMPP methods.
The details of the baselines are as follows:

\textbf{Methods without Pre-training.}
\begin{itemize}
    \item Siamese~\cite{koch2015siamese} utilizes dual convolutional neural networks to identify whether the input samples are from the same class. 
    \item ProtoNet~\cite{snell2017prototypical} classifies according to the inner-product similarity between the query sample and the prototype of each class.
    \item MAML~\cite{finn2017model} learns a model parameter initialization and adapts to new tasks via gradient descent.
    \item TPN~\cite{liu2019fewTPN} builds a relation graph and utilizes the entire query set for transductive inference.
    \item EGNN~\cite{kim2019edge} constructs a relation graph and predicts edge labels on the graph.
    \item IterRefLSTM~\cite{altae2017low} adapts a modification of Matching Networks for MPP tasks.
    \item PAR~\cite{wang2021property} employs a property-aware embedding function and adaptive relation graph learning to effectively propagate information among similar molecules.
    \item HSL-RG$^-$~\cite{ju2023few} leverages graph kernels and self-supervised learning to explore the structural semantics from both global-level and local-level granularities.
    \item GS-Meta~\cite{zhuang2023graph} constructs a molecule-property relation graph to leverage other available properties and reformulates an episode in meta-learning as a subgraph of this graph.
    \item ADKF-IFT~\cite{chen2022meta} is a framework for learning deep kernel Gaussian processes through interpolation between traditional deep kernel learning and meta-learning.
\end{itemize}

\textbf{Methods with Pre-training}
\begin{itemize}
    \item Pre-GNN~\cite{hu2020pretraining} is a pre-trained GNN using graph-level and node-level self-supervised tasks and is finetuned using a support set.
    \item Meta-MGNN~\cite{guo2021few} uses the Pre-GNN encoder and optimizes with self-supervised tasks in meta-training.
    \item Pre-PAR~\cite{wang2021property} ] is PAR initialized with Pre-GNN.
    \item HSL-RG~\cite{ju2023few} is HSL-RG$^-$ trained with the initialization of Pre-GNN.
    \item Pre-GS-Meta~\cite{zhuang2023graph} is the same as GS-Meta but uses the Pre-GNN encoder.
    \item Pre-ADKF-IFT~\cite{chen2022meta} is ADKF-IFT that begins with a pre-trained feature extractor.
    \item MolFeSCue~\cite{zhang2024molfescue} is a few-shot contrastive learning framework that combines the few-shot learning strategy with contrastive learning loss.
    \item PG-DERN~\cite{zhang2024property} is a few-shot learning model that introduces a dual-view encoder to learn a meaningful molecular representation by integrating information from node and subgraph.
\end{itemize}

\begin{table}[!t]
    \caption{Parameters in implementation.}
    \centering
    \renewcommand{\arraystretch}{1}
    \resizebox{0.6\columnwidth}{!}{
    \begin{tabular}{lr}
    \toprule[1.5pt] %添加表格头部粗线
    Description                                                                     & Value \\ 
    \midrule
    The dimension of embedding                                                      & 300\\
    The hidden size of classifier                                                   & 100-300  \\
    The hidden size of meta-training task sampler                                   & 300  \\
    The hidden size of auxiliary task sampler                                       & 300  \\
    The learning rate of the inner loop                                             & 0.01-1 \\
    The learning rate of the outer loop                                             & 0.001  \\
    The learning rate of meta-training task sampler                                 & 0.0005 \\
    The learning rate of auxiliary task sampler                                     & 0.0005  \\
    The dropout rate in molecular encoder                                           & 0.1-0.5 \\
    The dropout rate in the subgraph encoder                                        & 0.1-0.5 \\
    The contrastive loss weight in the outer loop                                   & 0.05  \\
    The temperature parameter of contrastive loss                                   & 0.05  \\
    \bottomrule[1.5pt] %添加表格底部粗线
    \end{tabular}
    }
    \label{tab: hyperparameters}
\end{table}

\begin{table*}[!t]
    \caption{The number of auxiliary tasks on Tox21, SIDER, MUV and PCBA dataset.}
    \centering
    \renewcommand{\arraystretch}{1}
    \resizebox{0.7\columnwidth}{!}{
    \begin{tabular}{ccccc}
    \toprule[1.5pt] %添加表格头部粗线
    Dataset & Tox21 & SIDER & MUV & PCBA \\
    \midrule
    The number of auxiliary tasks of 1-shot & 7 & 17 & 2 & 20 \\
    The number of auxiliary tasks of 10-shot & 7 & 17 & 2 & 20 \\
    \bottomrule[1.5pt] %添加表格底部粗线
    \end{tabular}
    }
    \label{tab: task_num}
\end{table*}

\begin{table*}[!t]
    \caption{The number of auxiliary tasks on ToxCast dataset.}
    \centering
    \renewcommand{\arraystretch}{1}
    \resizebox{1\columnwidth}{!}{
    \begin{tabular}{cccccccccc}
    \toprule[1.5pt] %添加表格头部粗线
    Sub-dataset & APR & ATG & BSK & CEETOX & CLD & NVS & OT & TOX21 & Tanguay \\
    \midrule
    The number of auxiliary tasks of 1-shot & 16 & 12 & 17 & 8 & 12 & 20 & 7 & 18 & 11 \\
    The number of auxiliary tasks of 10-shot & 16 & 12 & 17 & 8 & 12 & 20 & 7 & 18 & 11 \\
    \bottomrule[1.5pt] %添加表格底部粗线
    \end{tabular}
    }
    \label{tab: toxcast_task_num}
\end{table*}

\subsection{Implementation Details} \label{Implementation Details}
KRGTS is implemented in PyTorch~\cite{paszke2019pytorch} on a Ubuntu Server equipped with Intel(R) Core(TM) i7-8700K CPU, and 2 NVIDIA GeForce GTX 1080 Ti (with 11GB memory each). The experiments were conducted over 2000 training epochs based on Adam optimizer, with testing performed every 50 epochs, and repeated for 5 random runs on 2-way $K$-shot learning tasks, where $K\in\{1, 10\}$. Specifically, for the knowledge-enhanced molecule-property relation graph learning module, the following settings were used: (1) 9 and 1 top-$k$ molecule similarity preserved in subgraphs for 10-shot and 1-shot experiments, respectively; (2) a 5-layer GIN with a hidden size of 300 was employed as the graph-based molecular encoder; (3) the subgraph learning module consisted of a 2-layer GNN and a classifier comprising 2 layers MLP. For the meta-training task sampler, the task pool size $N_\text{pool}$ was set to 10, and the number of meta-training tasks sampled $N_\text{meta}$ was set to 5. The function $f_\text{meta}$ was implemented as a single-layer MLP. As for the auxiliary task sampler, $f_\text{auxi}$ was configured as a sequential module consisting of 2 layers MLP. Moreover, the rest parameters are listed in~\cref{tab: hyperparameters}. And the number of auxiliary tasks on each dataset is listed in~\cref{tab: task_num} and~\cref{tab: toxcast_task_num}.

\begin{table*}[!t]
    \caption{Performance on each sub-dataset of ToxCast in the 1-shot scenario.}
    \centering
    \renewcommand{\arraystretch}{1}
    \resizebox{1\columnwidth}{!}{
    \begin{tabular}{cccccccccc}
    \toprule[1.5pt] %添加表格头部粗线
    Method & APR & ATG & BSK & CEETOX & CLD & NVS & OT & TOX21 & Tanguay \\
    \midrule
    MAML & 64.59 & 55.45 & 60.36 & 61.02 & 66.22 & 59.84 & 62.15 & 59.52 & 60.92 \\
    ProtoNet & 57.08 & 54.92 & 53.92 & 60.25 & 66.25 & 54.87 & 63.11 & 58.27 & 58.32 \\
    EGNN & 67.06 & 57.28 & 60.82 & 60.10 & 71.53 & 56.56 & 66.08 & 63.32 & 74.80 \\
    PAR & 74.24 & 63.48 & 70.41 & 61.44 & 75.76 & 67.56 & 65.72 & 68.94 & 77.54 \\
    GS-Meta & \underline{87.90} & \underline{79.62} & \underline{85.94} & \underline{67.49} & \underline{78.16} & \underline{71.04} & \underline{72.36} & \underline{87.84} & \underline{89.97} \\
    KRGTS & \textbf{89.11} & \textbf{79.80} & \textbf{86.36} & \textbf{70.09} & \textbf{80.46} & \textbf{75.20} & \textbf{73.98} & \textbf{88.19} & \textbf{91.68} \\
    \midrule
    Pre-PAR & 84.69 & 70.38 & 79.89 & 66.57 & 77.83 & 72.51 & 70.41 & 80.33 & 86.64 \\
    Pre-GS-Meta & \textbf{89.49} & \textbf{81.69} & \textbf{87.28} & \underline{68.55} & \underline{78.66} & \underline{74.36} & \underline{73.56} & \underline{89.46} & \underline{91.10} \\
    Pre-KRGTS & \underline{89.45} & \underline{79.54} & \underline{86.84} & \textbf{72.50} & \textbf{81.21} & \textbf{76.63} & \textbf{73.68} & \textbf{89.51} & \textbf{92.15} \\
    \bottomrule[1.5pt] %添加表格底部粗线
    \end{tabular}
    }
    \label{tab: toxcast_1_shot}
\end{table*}

\begin{table*}[!t]
    \caption{Performance on each sub-dataset of ToxCast in the 10-shot scenario.}
    \centering
    \renewcommand{\arraystretch}{1}
    \resizebox{1\columnwidth}{!}{
    \begin{tabular}{cccccccccc}
    \toprule[1.5pt] %添加表格头部粗线
    Method & APR & ATG & BSK & CEETOX & CLD & NVS & OT & TOX21 & Tanguay \\
    \midrule
    MAML & 72.66 & 62.09 & 66.42 & 64.08 & 74.57 & 66.56 & 64.07 & 68.04 & 77.12 \\
    ProtoNet & 73.58 & 59.26 & 70.15 & 66.12 & 78.12 & 65.85 & 64.90 & 68.26 & 73.61 \\
    EGNN & 80.33 & 66.17 & 73.43 & 66.51 & 78.85 & 71.05 & 68.21 & 76.40 & 85.23 \\
    PAR & 82.74 & 68.86 & 74.65 & 67.76 & 78.33 & 70.79 & 69.12 & 77.34 & 83.39 \\
    GS-Meta & \underline{88.95} & \textbf{80.44} & \textbf{87.67} & \underline{69.50} & \underline{79.95} & \underline{74.77} & \underline{73.46} & \underline{88.78} & \underline{90.48} \\
    KRGTS & \textbf{89.95} & \underline{80.30} & \underline{87.22} & \textbf{72.15} & \textbf{81.25} & \textbf{76.84} & \textbf{74.63} & \textbf{89.19} & \textbf{91.96} \\
    \midrule
    Pre-PAR & 86.09 & 72.72 & 82.45 & 72.12 & 83.43 & 74.94 & 71.96 & 82.81 & 88.20 \\
    Pre-GS-Meta & \underline{90.15} & \textbf{82.54} & \textbf{88.21} & \underline{74.19} & \underline{86.34} & \underline{76.29} & \underline{74.47} & \textbf{90.63} & \underline{91.47} \\ 
    Pre-KRGTS & \textbf{90.31} & \underline{80.12} & \underline{87.92} & \textbf{76.63} & \textbf{86.97} & \textbf{77.52} & \textbf{75.11} & \underline{89.83} & \textbf{91.73} \\
    \bottomrule[1.5pt] %添加表格底部粗线
    \end{tabular}
    }
    \label{tab: toxcast_10_shot}
\end{table*}

\subsection{Performance of ToxCast Sub-datasets}
Consistent with~\cite{zhuang2023graph}, each sub-dataset was further divided into meta-training and testing sets, as illustrated in~\cref{tab: toxcast_dataset}. The results of the sub-datasets are presented in \cref{tab: toxcast_1_shot} and~\cref{tab: toxcast_10_shot}. Also, the results in~\cref{tab: result} and~\cref{tab: pre-result} are averaged from~\cref{tab: toxcast_1_shot} and~\cref{tab: toxcast_10_shot}. From the experimental results, it can be observed that KRGTS shows significant gains on most sub-datasets, leading to an overall performance improvement on the entire dataset, further demonstrating the effectiveness of KRGTS.
%从实验结果观察可得，KRGTS在多数子数据集上均有明显增益，特别地，在Tanguay上达到了92.15的性能，实现进而实现了整个数据集性能得提升，进一步证明了KRGTS的有效性。

\subsection{Computation Resource Analysis}
\begin{wraptable}{r}{6.9cm}
    \caption{Time complexity comparison.}
    \centering
    \resizebox{0.5\textwidth}{!}{
    \begin{tabular}{ccc}
        \toprule[1.5pt] %添加表格头部粗线
        Dataset & GS-Meta & KRGTS\\
        \midrule
        Tox21 & 85.85\%(2.37s, 6030MB) & \textbf{87.19\%}(3.10s, 3134MB)\\
        SIDER & 83.72\%(2.45s, 7674MB) & \textbf{84.83\%}(5.23s, 4780MB)\\
        MUV & 67.11\%(2.47s, 6302MB) & \textbf{72.63\%}(1.63s, 4598MB)\\
        \bottomrule[1.5pt] %添加表格底部粗线
    \end{tabular}
    }
\label{tab: complexity}
\end{wraptable}
%为做全面评估，我们对比了KRGTS和GS-Meta的时间复杂度。具体来说，由于KRGTS的运行时间主要受采样器的任务表征学习影响，我们选择了三个标签稀疏性不同的数据集，并记录了它们每一个epoch的训练时间和显存消耗。实验结果如表三所示。从整体来看，KRGTS在性能和显存消耗上都远优于GS-Meta，并且当标签较为稀疏时其在预测性能和运行时间上都具有显著优势。此外，可以观察到KRGTS的运行时间会受到候选辅助任务数量的影响。其中，MUV数据集的标签最为稀疏，运行时间也最快。而Tox21和SIDER相对来说更加稠密，因此时间复杂度也有相应增加，但这一问题是可以通过多进程训练机制来缓解的。
To conduct a comprehensive performance evaluation, we compared the time complexity of KRGTS with GS-Meta. Specifically, considering that the runtime of KRGTS is primarily influenced by subgraph learning in the sampler, we selected three datasets (MUV, Tox21, SIDER) with different scales and recorded the training time and memory consumption for each epoch on each dataset. The experimental results are presented in Table 3. Overall, KRGTS significantly outperforms GS-Meta in terms of performance and memory consumption, especially in cases of sparse labels, where it shows impressive superiority in performance and runtime. Additionally, we observed that the runtime of KRGTS is affected by the number of candidate auxiliary tasks. For instance, KRGTS exhibits the fastest runtime on the MUV dataset, which has the sparsest labels. Although the time complexity increases on the Tox21 and SIDER datasets due to their relatively denser labels, this issue can be effectively mitigated by employing a multi-process training mechanism.

\end{document}